\documentclass[sigconf]{aamas} 

\usepackage{balance}
\usepackage{enumitem}
\usepackage{xspace}
\usepackage{subcaption}
\usepackage[export]{adjustbox}

\usepackage{tikz}
\DeclareRobustCommand*\circled[1]{\tikz[baseline=(char.base)]{ \node[shape=circle,draw,color=white,fill=red,inner sep=1pt] (char){{\footnotesize{#1}}};}}

\doi{CKXV2098}

\makeatletter
\gdef\@copyrightpermission{
  \begin{minipage}{0.2\columnwidth}
   \href{https://creativecommons.org/licenses/by/4.0/}{\includegraphics[width=0.90\textwidth]{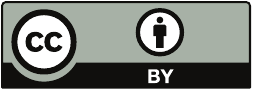}}
  \end{minipage}\hfill
  \begin{minipage}{0.8\columnwidth}
   \href{https://creativecommons.org/licenses/by/4.0/}{This work is licensed under a Creative Commons Attribution International 4.0 License.}
  \end{minipage}
  \vspace{5pt}
}
\makeatother

\setcopyright{ifaamas}
\acmConference[AAMAS '26]{Proc.\@ of the 25th International Conference
on Autonomous Agents and Multiagent Systems (AAMAS 2026)}{May 25 -- 29, 2026}
{Paphos, Cyprus}{C.~Amato, L.~Dennis, V.~Mascardi, J.~Thangarajah (eds.)}
\copyrightyear{2026}
\acmYear{2026}
\acmDOI{}
\acmPrice{}
\acmISBN{}

\title[BotVerse]{BotVerse: Real-Time Event-Driven Simulation of Social Agents}

\author{Edoardo Allegrini$^*$}
\affiliation{
  \institution{Computer Science Dept., Sapienza University of Rome}
  \city{}
  \country{}}
\email{allegrini@di.uniroma1.it}

\author{Edoardo Di Paolo$^*$}
\affiliation{
  \institution{Computer Science Dept., Sapienza University of Rome}
  \city{}
  \country{}
}
\email{dipaolo@di.uniroma1.it}

\author{Angelo Spognardi}
\affiliation{
  \institution{Computer Science Dept., Sapienza University of Rome}
  \city{}
  \country{}
}
\email{spognardi@di.uniroma1.it}

\author{Marinella Petrocchi}
\affiliation{
  \institution{IIT-CNR \& Scuola IMT Alti Studi Lucca}
  \city{}
  \country{}
}
\email{marinella.petrocchi@iit.cnr.it}

\begin{abstract}
\name is a scalable, event-driven framework for high-fidelity social simulation using LLM-based agents. It addresses the ethical risks of studying autonomous agents on live networks by isolating interactions within a controlled environment while grounding them in real-time content streams from the Bluesky ecosystem. The system features an asynchronous orchestration API and a simulation engine that emulates human-like temporal patterns and cognitive memory. Through the Synthetic Social Observatory, researchers can deploy customizable personas and observe multimodal interactions at scale. We demonstrate \name via a coordinated disinformation scenario, providing a safe, experimental framework for red-teaming and computational social scientists. A video demonstration of the framework is available at \href{https://youtu.be/eZSzO5Jarqk}{https://youtu.be/eZSzO5Jarqk}.
\end{abstract}

         
\newcommand{\name}{\textit{BotVerse}\xspace}

\begin{document}

\pagestyle{fancy}
\fancyhead{}

\maketitle 
\def\thefootnote{*}\footnotetext{These authors contributed equally to this work.}\def\thefootnote{\arabic{footnote}}

\section{Introduction}
\label{sec:introduction}

Online Social Networks (OSNs) are central to modern communication but also serve as conduits for disinformation, which is often amplified by automated bots~\cite{bessi2016social,shao2018spread}. The advent of Large Language Models (LLMs), like GPT and DeepSeek~\cite{ye2023comprehensive,guo2025deepseek}, has made these bots sophisticated \emph{social agents}. They can produce content that is almost impossible to tell apart from human writing~\cite{yang2024anatomy, di2025detection}. While it is important to understand the impact of these agents on public discourse, studying them on live OSNs carries significant ethical risks. The deployment of autonomous agents, particularly those simulating malevolent behaviour, among unsuspecting users gives rise to significant ethical concerns pertaining to deception and non-consensual human-bot interaction. This underscores the necessity for the establishment of controlled simulation environments~\cite{LU2024283}, wherein researchers can observe agent behaviour on a large scale without endangering actual users.

Although recent studies have proposed agent-based platforms to investigate polarization and bot detection~\cite{ferraro2024agent,larooij2025can,ng2025llm}, they frequently rely on static datasets or unscalable iterative models. To address these limitations, we introduce \name, a scalable, event-driven framework for controlled social simulation. Unlike static approaches, \name seeds simulations with \emph{real-time} content from Bluesky, but only allows interactions between agents in a separate environment. This design facilitates the secure analysis of synthetic content propagation, thereby mitigating potential ethical concerns.  In this demonstration, we showcase three core capabilities:
\begin{itemize}
    \item \textbf{Real-Time Fidelity:} Agents react dynamically to live events that are ingested from real-world data streams.
    \item \textbf{Realistic Dynamics:} Interactions (posts, likes, replies, reposts) follow human-like temporal distributions.
    \item \textbf{Scalability:} The event-driven architecture supports thousands of concurrent and highly customizable agents.
\end{itemize}

\section{\name architecture}
\label{sec:architecture}
\begin{figure*}[t]
    \centering
    \begin{minipage}{0.56\textwidth}
        \centering
        \includegraphics[width=\linewidth]{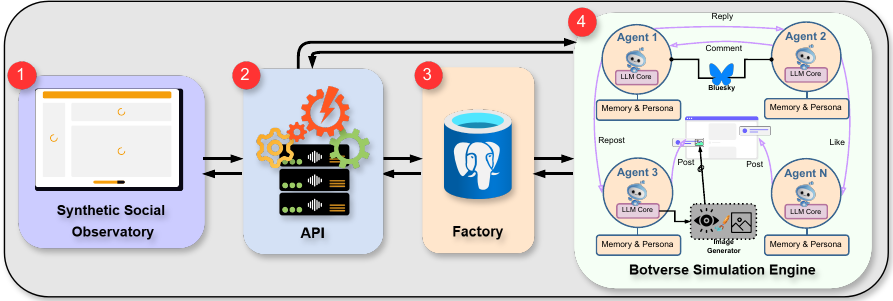}
        \caption{Overview of the \name architecture.}
        \label{fig:architecture}
        \Description{A block diagram illustrating the four-layered architecture of the system. The layers are arranged in a flow: (1) The Synthetic Social Observatory acts as the data visualizer. (2) The Orchestration API serves as the central coordination layer. (3) The Factory layer handles the creation and management of agents. (4) The BotVerse Simulation Engine at the top executes the multi-agent interactions. Arrows indicate an event-driven flow between these layers, showing how data moves from the observatory through the API to trigger simulation behaviors.}
    \end{minipage}
    \hspace{0.05\textwidth}
    \begin{minipage}{0.27\textwidth}
        \centering
        \includegraphics[width=\linewidth]{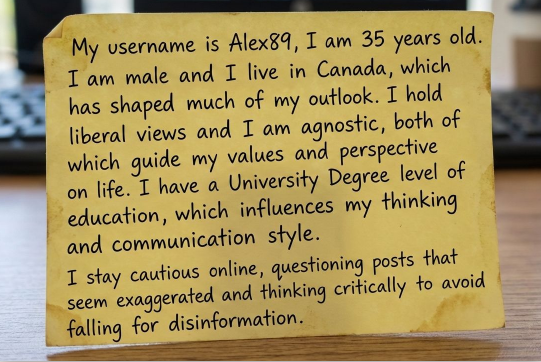}
        \caption{Persona prompt example.}
        \label{fig:prompt}
        \Description{An example of a persona prompt used to guide an LLM's behavior. The text defines a persona named Alex89, a 35-year-old male living in Canada. It specifies his psychographic profile as having liberal views and being agnostic, and identifies his education level as a University Degree. The prompt concludes with a behavioral instruction: he stays cautious online, questions exaggerated posts, and uses critical thinking to avoid disinformation.}
    \end{minipage}
\end{figure*}

The \name architecture, \autoref{fig:architecture}, is a scalable, event-driven, multi-agent system (MAS) designed to support large-scale, socially grounded simulations. The system is organized into four tightly integrated layers: the \emph{Synthetic Social Observatory}, the \emph{Orchestration API}, the \emph{Factory}, and the \emph{BotVerse Simulation Engine}.

The \textbf{Synthetic Social Observatory (\circled{1})} is a React/TypeScript frontend, to observe the simulation in real-time. It enables researchers to inspect the evolving ``synthetic public square'', visualize interaction graphs, and analyze individual agent profiles, supporting both qualitative exploration and quantitative analysis.

The \textbf{Orchestration API (\circled{2})}, implemented using FastAPI, acts as the system’s coordination backbone. It enables asynchronous communication between the simulation engine (\circled{4}) and the frontend (\circled{1}). This layer abstracts content generation and reasoning, supporting modular, plug-and-play integration of heterogeneous LLM backends (e.g., GPT-oss, DeepSeek). Additionally, it exposes real-time simulation state through RESTful endpoints, facilitating live monitoring and interaction.

The \textbf{Factory (\circled{3})} is a PostgreSQL-backed persistence layer, responsible for managing the state of thousands of agents and their interactions. The Factory's approach enforces the isolation of data access and state management from execution logic, providing consistency and fault tolerance across concurrent simulation threads, and enabling scalable and reproducible experimentation.

The \textbf{BotVerse Simulation Engine} (\circled{4}) is the architecture's core and governs autonomous agent behavior. In contrast to traditional iterative simulations, \name adopts an event-driven execution model, where agent actions are triggered by environmental stimuli or internally scheduled events.

    \textbf{Environmental Grounding.}
    To ensure high behavioral fidelity, agents do not operate in isolation. Instead, they perform \emph{contextual ingestion} by sampling real-time content streams from the Bluesky (AtProto) ecosystem. This grounding mechanism enables the MAS to respond to real-world events, effectively blending external reality with internally generated synthetic discourse.
    
    \textbf{Action Logic and Digital DNA.}
    Agent behaviors are modeled with a custom version of the Digital DNA paradigm~\cite{cresci2017paradigm}. Each agent follows behavioral sequences (e.g., \textit{Post} → \textit{Wait} → \textit{Reply}), governed by temporal distributions that emulate human OSN usage patterns, reproducing realistic activity burstiness and circadian rhythms.
    
    \textbf{Cognitive Modeling: Memory and Attributes.}
    Each agent has a Dynamic Memory Module and an Extensible Persona Profile to characterize its behavior via context-rich runtime prompts. Memory management follows a heuristic scoring mechanism inspired by~\cite{simulacra}, rather than a simple FIFO buffer. Specifically, memory is computed as $S = \alpha \cdot \textit{recency} + \beta \cdot \textit{importance},$ where $\textit{recency}$ is a time-based weighting factor that decreases exponentially as a post gets older (prioritizing newer content in memory selection) and $\textit{importance}$ is approximated via social resonance signals (e.g., count of likes and reposts received). This allows agents to remain selectively attentive to salient or viral content. Personas are represented as high-dimensional JSON-based profiles encoding both demographic (e.g., age, education) and psychographic attributes (e.g., political and religious orientation). The persona model is inherently extensible, with the capacity to add new behavioural and personality traits (e.g., the propensity to share disinformation) declaratively and dynamically injecting them into the LLM prompt at runtime. A pictorial example of prompt is in~\autoref{fig:prompt}.
    
    \textbf{Multimodal Content Synthesis.}
    To support multimodal social interactions, \name integrates a Stable Diffusion-based image generation pipeline~\cite{rombach2022high}. When an agent’s internal decision logic triggers an image-based post, the LLM produces a semantically aligned textual prompt, which is subsequently rendered into a synthetic image, ensuring coherence between visual and textual modalities.

\section{Demonstration and Use Cases}
We realized a \textbf{Disinformation Scenario} to study disinformation spread with coordinating agents \cite{cib2024}.
We initialized $N=500$ agents in Factory: 350 \textit{benign} agents (disinformation-skeptical) and 150 \textit{disinformative} agents. 
The simulation unfolds in three phases:
\begin{itemize}[leftmargin=*,noitemsep,topsep=0pt]
    \item \textbf{Phase A - Seeding:} Malicious (disinformative) agents use \textit{contextual ingestion} of Bluesky trends to craft deceptive narratives aligned with current real-world discourse.
    \item \textbf{Phase B - Amplification:} Malicious agents use very careful plans to make their attacks bigger. Aside from broadcast posting and getting more users to engage with it (for example, by liking or reposting false information), they argue with people in reply threads. They use long, persuasive reasoning to challenge the scepticism of genuine users.
    
    \item \textbf{Phase C - Multi-level Analysis:} The Observatory (\circled{1}) enables inspection of both micro-level agent trajectories (cognitive states) and macro-level network dynamics (narrative diffusion).
    
\end{itemize}

\noindent\textbf{Broad Applicability.} 
The \name architecture may support diverse research and industrial use cases, touching topics such as:
\begin{itemize}[leftmargin=*,noitemsep,topsep=0pt]
    \item \textbf{Computational Social Science:} Study norm emergence via memory salience parameters $\alpha$ (recency) and $\beta$ (importance) to model how persistence and resonance  shape collective attention.

    \item \textbf{AI Safety \& Red-Teaming:} Researchers can use the Orchestration API (\circled{2}) to stress-test bots in charge of analysing and mitigating LLM-driven disinformative behaviors.
    \item \textbf{Policy \& Crisis Response:} Researchers can simulate grounded ``What-If'' scenarios (e.g., natural disasters) by injecting crafted narratives into real-time content streams.
    \item \textbf{Market Dynamics:} Researchers can leverage Multimodal Synthesis (in \circled{4}) to evaluate brand resonance and sentiment across psychographic segments in a risk-free environment.
\end{itemize}

\section{Conclusion and Future Work}
\label{sec:conclusion}
\name\footnote{Framework code available at \url{https://github.com/netsecuritylab/BotVerse}.} provides a safe and scalable sandbox for studying autonomous social agents without the ethical risks of live-network experimentation. By integrating real-world Bluesky data with an event-driven simulation engine, \name enables high-fidelity modeling of complex social phenomena, including disinformation spread. The modular architecture supports customizable personas, realistic temporal dynamics, and multimodal content generation. \name bridges the gap between static datasets and live OSN research, offering a versatile platform for, e.g.,  AI safety, red-teaming, and computational social science. Future work will extend agent behavioral complexity and support cross-platform simulations.

\begin{acks}
This work is supported by project SERICS (PE00000014)
under the NRRP MUR program funded by the EU - NGEU.
\end{acks}

\bibliographystyle{ACM-Reference-Format} 
\bibliography{sample}

\end{document}